\documentclass{rspublic}
\usepackage{geometry}                
\usepackage{graphicx}
\usepackage{amssymb}
\usepackage{epstopdf}

\begin{document} 
\title[MOND and TeVeS]{Tensor-Vector-Scalar modified gravity: from small scale to cosmology} 
\author[J. Bekenstein]{Jacob D. Bekenstein} 
\affiliation{Racah Institute of Physics, The Hebrew University of Jerusalem, Givat Ram, Jerusalem, 91904, Israel} 
\label{firstpage} 
\maketitle 
\begin{abstract}{Dark Matter,galaxies,gravitational lensing,MOND,TeVeS} 
The impressive success of the standard cosmological model has suggested to many that its ingredients are all one needs to explain galaxies and their systems.  I summarize a number of known problems with this program.  They might signal the failure of standard gravity theory on galaxy scales. The requisite hints as to the alternative gravity theory may lie with the MOND paradigm which has proved an effective summary of galaxy phenomenology.  A simple nonlinear modified gravity theory does justice to MOND at the nonrelativistic level, but cannot be consistently promoted to relativistic status. The obstacles were first sidestepped with the formulation of T$e$V$e\,$S, a covariant modified gravity theory.  I review its structure, its MOND and Newtonian limits, and its performance in face of galaxy phenomenology.  I also summarize features of T$e$V$e\,$S cosmology and describe the confrontation with data from strong and weak gravitational lensing.
\end{abstract}

\section{From standard cosmology to small scale structure}
\label{ref:intro}

The agreement between the measured light elements cosmic abundances and predictions by the theory of big bang nucleosynthesis and that between the measured spectrum of fluctuations in the cosmic microwave background (CMB) and predictions from the general relativistic theory of growth of cosmological perturbations have convinced many that the standard concordance cosmological model must be near the mark.    To what extent are the expectations from this model for the issue of small scale structure confirmed?

For my purposes here small scale structure refers to systems ranging from the scale of large clusters of galaxies down to those of dwarf galaxies.  According to the standard model, all these originated from initially small fluctuations in the dark matter (DM) density distribution which grew prior to recombination.  As the baryonic plasma became neutralized it began to fall into gravitational potential wells generated by the earlier collapsing DM.  This was to give rise to all the small structure we see around us, each luminous galaxy embedded in a heavier halo of DM. 

In fact, well before the standard cosmological model was in place, Ostriker \& Peebles~[1] realized that in Newtonian gravity a pure disk galaxy supported solely by rotation is unstable to the formation of a bar.  In view of the relative paucity of barred galaxies, they proposed that the typical disk galaxy nestles in the gravitational potential of a stabilizing massive encompassing invisible halo supported by random motion.  This scenario was applied to explain the fact, just then becoming established, that rotation curves (RCs) of disk galaxies are quite flat outside the central regions.  A flat rotation curve follows if a halo with $r^{-2}$ density profile dominates the gravitational field outside the central regions.   Modern dark halo modelling, inspired by the Navarro-Frenk-White DM halo density profile (NFW)~[2] from cosmological simulations, has been widely acclaimed as explaining the shapes of the RCs.

In the typical RC the passage from the central rising stage (dominated by the baryonic disk) to the flat part (dominated by the halo) is featureless. Bahcall \& Cassertano~[3] realized that this requires the parameters of disk and halo to be fine tuned together.   A more dramatic example of fine tuning is posed by the Tully-Fisher (TF) relation for disks.  In its modern form by McGaugh~[4] the baryonic mass in the galaxy is accurately proportional to the fourth power of the rotation velocity (in the flat part), with a proportionality coefficient independent of galaxy type.  Since the rotation velocity is dominated by the DM halo, the halo parameters must be fine tuned to the mass of the baryonic disk.  But fine-tuning begs the question; can DM halo modelling get us out of it?

The NFW halo's mass density profile starts like $r^{-1}$ near the centre, and then steepens to $r^{-3}$ in the outer parts.  Only in the intermediate region is it like the $r^{-2}$ profile that gives a flat RC; and indeed the models based on NFW do not well reproduce the observed flatness of RCs well beyond the optical edge of the galaxies.  Further, in NFW halos the DM mass scales as the velocity dispersion cubed, and virial arguments tie the RC amplitude to the DM velocity dispersion.  How to get the $M\propto v^4$ TF law?    It has been a pious hope of galaxy DM proponents that in the course of infall of baryonic gas into a DM halo potential well, gas dynamical processes will fine tune disk and halo together to give the TF relation.  This fine tuning should also---so the hope---erase any incipient feature at the joint between the disk dominated rising part of the RC and the halo dominated flat part.  But the TF correlation is among the tightest in the whole of astrophysics; in fact, it is about as narrow as allowed by the \emph{observational} errors.  As Bob Sanders emphasized to me, turbulent gas dynamics would naturally lead to a spread in the amount of gas mass falling into each of a set of identical halo potential wells.  This will sabotage the process attempting to fine tune, even if it exists.

The  central $r^{-1}$ mass density profile of the NFW halo means there is a mass cusp at the centre.  In low surface brightness galaxies (LSBs) the baryon density (mostly gaseous) is so low that the halo should dominate dynamics everywhere.  But careful study of extant RCs by Chen \& McGaugh~[5] does not disclose the fast motions near the centres that would be expected in the presence of cusps.   Again, one must rely on the hope that gas dynamics feedback on the halo have flattened the cusps. 

Cosmological simulations predict that NFW halos form in a hierarchy, with small halos arising together with each big one.  If each halo eventually cuddles a galaxy, with galaxy baryonic mass increasing monotonically with halo mass, there end up being many more small satellite galaxies per big galaxy than observed in our neighbourhood (galaxy surveys are now showing this is also true elsewhere).   This embarrassment could again be resolved if gas dynamics is such that the small halos remain unpopulated by baryons and thus invisible.  Such eventuality could be checked with gravitational lensing which is also sensitive to sterile halos.   (Claims of detection of the corresponding small scale substructure around big galaxies have not been confirmed.)

\section{The MOND paradigm}
\label{sec:paradigm}

Thus, while the standard cosmological model may be convincing on the grounds of its predictive successes, as we follow its logical implications towards smaller scales, we meet with embarrassments.   Of course all these could be the product of astrophysical complications not yet accounted for.  But, logically, the possibility cannot be excluded that standard gravity theory, which is implicitly assumed in all the above deductions, is failing on small scales.

As the ubiquity of the flat RCc and the wide applicability of the TF luminosity vs rotational velocity were being internalized, Milgrom~[6] realized that the naive Newtonian description of a galaxy containing only baryonic matter fails in a way that correlates with the system's  typical acceleration (in units of a certain scale $a_0$), rather than with its length or mass scales.  (There is a beautiful pair of graphs in McGaugh \& Sanders' comprehensive review~[7]  that illustrates this with modern data.)   While Milgrom was not the first to suggest a departure from Newtonian dynamics as the resolution of the above problems, his phenomenological scheme---MOND---has been by far the most fruitful in this respect.  

MOND posits~[6,8] that in astronomical systems the acceleration $\mathbf{a}$ is related to the Newtonian gravity field $\mathbf{g}_N$ sourced by baryonic matter by
\begin{equation}
\tilde\mu(|\mathbf{a}|/a_0) \mathbf{a} = \mathbf{g}_N.
\label{MOND}
\end{equation}
Here $\tilde\mu$ is a function of positive argument interpolating between 0 at low argument and 1 at large argument (both compared to unity), for example, the ``simple'' interpolation function
\begin{equation}
\tilde\mu(x) = \frac{x}{1+x}.
\label{simple}
\end{equation}
Applied to test particles in an ambient gravitational field, equation~(\ref{MOND})  can be used to predict rotation curves from the observed baryonic mass distribution.   It is easily seen that this relation forces the RC to become flat in a disk's outskirts, where $a\ll a_0$ (postdiction of observed systematics) and also requires the baryonic total mass $M$ (source of $\textbf{g}_N$) to scale as $v_c^4$ where $v_c$ is the rotation velocity in the flat part of the RC (prediction of the modern form of the TF law).

Other early predictions of MOND in this simple form are~[7,8]: 

\begin{itemize}
\item A LSB (defined as a disk whose \emph{surface} mass density is everywhere below $G a_0$) should have an especially large mass discrepancy (ratio of dynamically inferred mass to visible baryonic mass).   When years later the first LSB RCs were measured, this prediction was corroborated.  

\item A LSB's  RC must always rise, in contrast to a higher surface density disk whose RC rapidly peaks and then descends to a plateau.  The predicted detailed shape of the LSB's  RC is independent of the choice of $\tilde\mu$ function.   The rising behaviour of the RC is well documented today, and MOND predictions of LSB RC shapes are especially successful.  

\item As a rule globular clusters must be well described Newtonially in terms of visible stellar mass because their internal Newtonian fields exceed $a_0$.  This is now well established; yet at the time of the prediction it was unexpected that globulars should have no DM halos (even today it is not clear why they do not).  

\item Dwarf spheroidal galaxies (sparse assemblies supported by random stellar motion) should have big mass discrepancies because they all have surface mass densities well below $G a_0$.   The large dynamical mass-to-light ratios of dwarf ellipticals are well documented today.

\item  MOND can predict the presence of features in the RC of a disk with normal surface mass density in terms of the disk's detailed baryonic mass density profile, provided one uses the ``correct'' $\tilde\mu$.  The trend today is to use equation~(\ref{simple}).  Over 100 fairly well measured RCs of such galaxies have been fit by MOND, and as a rule these fits are as good as any DM halo fit.  In contrast to the classic one-parametric NFW halo family,  models used in DM halo fits~[9] have two parameters, the halo's core radius (to excise the embarrassing cusp), and its central velocity dispersion.  To these is added the mass-to-light ratio $\Upsilon_\star$ of the galaxy's stellar component.  MOND is more parsimonious, requiring only  one parameter, $\Upsilon_\star$; yet it gives fits as good or better than do the DM models.
\end{itemize}

Nobody is prefect, and MOND is no exception.  While it gives a better account of the velocity dispersions in clusters of galaxies than does Newtonian theory \emph{without} DM, it still requires the addition of unseen material~[7].  Ironically, even this failure led to a good early prediction; Milgrom suggested that the problem results from the existence, in all clusters, of large quantities of gas  which had not yet been seen~[10].  In the mid 1980's such X-ray emitting gas was indeed found in clusters, to the tune of five times the mass in galaxies or so. While this finding tends to help, the gas is not enough to wholly vindicate MOND, which still must rely on an amount of invisible matter comparable to that of the gas mass. 

Another partial success of MOND has to do with giant and dwarf ellipticals.  Observationally these fall on a ``fundamental plane'' FP in the space of central velocity dispersion, effective radius and mean surface brightness. The locus predicted by Newtonian theory for virially equilibrated galaxies with constant $\Upsilon_*$ is tilted with respect to the observed FP.  This can be fixed by appealing to a DM halo as well as strong velocity anisotropy.  According to a recent detailed study by Cardone et al. MOND does better but leaves some tilt too~[11].  Possibly taking into account the external field effect (see \S\ref{ref:gravity}) of the nearby galaxies (ellipticals are mostly found in densely populated regions) will mend the problem.

Mild failures aside, it is clear that there is a broad range of masses, $10^6-10^{11} M_\odot$, in which systems adhere to MOND in their systematics.  This must be telling us something; logically there are the following possibilities.
a) MOND is merely an efficient summary of the way DM is distributed in the said systems? 
b) MOND reveals the dependence of inertia on acceleration for small accelerations?  
c) MOND betrays hitherto unknown forces particularly effective at astronomical scales?  
d) MOND encapsulates departures from standard Newtonian-Einsteinian gravity theory at the mentioned scales?  

Possibility a) has to contend with the fact, mentioned above, that  predictions from dynamical simulation for DM halos (NFW)  run into observational problems (e.g. cusps, paucity of satellites).   The hope that backreaction of the gas dynamics on the halos will resolve all problems seems to me to be a wishful hope.   The best theory for option b) is a nonlocal theory of inertia which is extremely hard to calculate with~[12].  One is stymied in this direction.  Option c) alludes to the ``fifth force'' paradigm.  Were the fifth force weaker than Newtonian gravity by an order of magnitude, it would become irrelevant for our galactic scale problem.  Were it stronger by an order of magnitude, it would swamp gravity and contravene the facts.  Such fifth force must couple only to mass, just as does gravity, in order to preserve the universality of free fall which we seem to see working in galactic orbits.  Thus we may just as well think of this force as a facet of gravity.  Possibility d) is my topic here.  My talk is an update of several reviews, which should be consulted for details and a fuller list of references~[13-15].

\section{MOND as gravity}
\label{ref:gravity}
Formula (\ref{MOND}) is good for test particles in an external field.  Applied to a composite system, e.g. a binary galaxy, it entails nonconservation of momentum, energy and angular momentum (because the inertia of each particle can depend on its acceleration).  A consistent embodiment of the MOND paradigm as a gravity theory must be based on a Lagrangian to enforce the conservation laws.  In fact, if  in the usual  Newtonian Lagrangian density we make  the replacement
\begin{equation}
|\nabla\Phi_N|^2 \Longrightarrow a_0^2\, F(|\nabla\Phi|^2/a_0^2)
\label{replace}
\end{equation}
 will do the trick; here $\nabla\Phi_N=-\mathbf{g}_N$ while $\Phi$ is the true gravitational potential ($\mathbf{a}=-\nabla\Phi$) that couples to mass density $\rho$ just as $\Phi_N$ does in the Newtonian theory.  The Euler-Lagrange procedure leads to the modified Poisson equation~[16]:
 \begin{equation}
\label{mod-Poisson}
\nabla\cdot[\tilde\mu(|\nabla\Phi|/a_0) \nabla\Phi]=-8\pi G\rho;\qquad \tilde \mu(\surd  x)\equiv dF(x)/dx. 
\end{equation}
Here $\tilde\mu(x)$ is identified with that in equation (\ref{MOND}).
This theory is called AQUAL (for AQUAdratic Lagrangian); it respects the usual conservation laws.

For spherical, cylindrical and planar geometry, the solution of equation~(\ref{mod-Poisson}) together with $\mathbf{a}=-\nabla\Phi$ gives formula~(\ref{MOND}): AQUAL is a decent nonrelativistic theory  embodying the MOND paradigm.  For lower symmetry the corrections are important, and have been studied somewhat with numerical codes.  Among the additional insights issuing from AQUAL I mention
\begin{itemize}
\item AQUAL explains why an internally high acceleration object, e.g. a star, orbits in a galaxy according to the MOND dynamics that pertain to the low acceleration of the orbit of its centre of mass~[16].
\item AQUAL confirms the suspected existence of the \emph{external field effect}: an internally weak acceleration object, e.g. an open cluster, subject to an external field  $>a_0$ has nearly Newtonian dynamics~[6,16].
\item AQUAL allows study of dynamical evolution of multiple systems, e.g. binary galaxy~[17].
\item  AQUAL gives a systematic account of regularities in pressure supported systems, e.g. early type galaxies~[18].
\end{itemize}

These successes support the notion that MOND encapsulates some deviation of gravitation physics from the standard Newton-Einstein paradigm.   Of course one would also like a relativistic version of the theory to confront data from gravitational lenses, cosmology, etc.   In AQUAL we managed to build the Lagrangian because the fundamental quantity $\nabla\Phi$ has the meaning and dimension of the acceleration field, and can thus be compared with $a_0$ to delineate the MOND dominant region.   In the geometric (Einstein) formulation of gravity theory, the analog of $\nabla\Phi$ is the connection or Christoffel symbol $\Gamma^\alpha_{\beta\gamma}$ associated with the metric.  But this last is not a tensor, and so cannot be used undifferentiated to form a covariant Lagrangian.  Worse, by judicious choice of coordinates one can make $\Gamma^\alpha_{\beta\gamma}=0$ locally, but obviously the MOND regime should not be delineated by  the choice of coordinates!  We really need a different acceleration field.

The simplest thing to do is to replace  AQUAL's  $\Phi$ by a relativistic scalar field $\psi$ bearing the same dimensions, those of $c^2$.  We keep the metric $g_{\alpha\beta}$ of general relativity (GR) governed by the usual action.  But we break with GR by requiring that matter ``live'' in the metric $\tilde g_{\alpha\beta}= \exp(2\psi/c^2)\,  g_{\alpha\beta}$  conformally related to $g_{\alpha\beta}$.   The Lagrangian for $\psi$ is a covariant version of that for $\Phi$ in AQUAL; the corresponding equation is a wave equation version of equation~(\ref{mod-Poisson}) with $\rho$ replaced by minus the trace of the energy-momentum tensor of the matter, $T$.   All this characterizes  R(elativistic)AQUAL~[16].  Analysis of nonrelativistic motion identifies the potential $\Phi$ with $\Phi_N+\psi$. For quasistatic nonrelativistic systems one can ignore time derivatives as well as replace $T$ by $-\rho$; then the equation determining $\psi$ is essentially equation~(\ref{mod-Poisson}).   Thus in the regime where $|\nabla\psi| c^{-2}/a_0\ll1$,  $|\nabla\psi|\gg  |\nabla\Phi_N|$ and one can take  $\Phi\approx \psi$ and so recover AQUAL. RAQUAL thus inherits all good properties of AQUAL and reproduces MOND.   But two serious deficiencies of RAQUAL were  apparent early~[16]. 
 
\begin{itemize}
\item Small perturbations of $\psi$ about a background solution propagate superluminally in regions with low acceleration.
\item The gravitational deflection of light in RAQUAL is about the same as that in GR for the same mass.
\end{itemize}

Superluminality raises the spectre of acausal behaviour; it is worrying, but perhaps not disabling  since it does not run in the face of any known observations.  However, already by the early 1980's it was clear that in clusters of galaxies which happen to gravitationally lense quasars, the deflection of light is much larger than anticipated by GR models with no DM included.  Thus RAQUAL, which as a matter of principle eschews DM, runs counter to the facts.  Wherein lies the problem?  First, the energy momentum tensor of $\psi$ itself is quite negligible compared to that of the matter, so the metric $g_{\alpha\beta}$ is about the same as in GR.  Second, the metric $\tilde g_{\alpha\beta}$, to which electromagnetic fields couple, is conformal to  $g_{\alpha\beta}$, and it is well known that conformally related metrics are totally equivalent with respect to electrodynamics.  The solution to the problem is thus obvious: make the passage from $g_{\alpha\beta}$ to $\tilde g_{\alpha\beta}$ nonconformal.  For example, the two metrics could be related \emph{disformally},  $\tilde g_{\alpha\beta}=\exp(2\psi/c^2)(g_{\alpha\beta}+\varpi\psi_{,\alpha}\psi_{,\beta})$, with $\varpi$ a function of the invariant norm of $\psi_{,\alpha}$~[19].  The requirements that $\tilde g_{\alpha\beta}$ possess an inverse, that the null cone of $g_{\alpha\beta}$ not lie outside that of $\tilde g_{\alpha\beta}$ (no superluminal propagation of gravitons) and that $\psi$ bear positive energy density force $\varpi<0$.  Calculation then shows that light deflection is actually a little weaker in the new theory than in RAQUAL~[20].  So purely scalar-tensor scalar theories fail to cope with the facts of life.

What the scalar field cannot accomplish alone, it may accomplish with help from a vector field.  Sanders proposed the disformal relation (from now on I set $c=1$)
\begin{equation}
\label{gg}
\tilde g_{\alpha\beta} = e^{-2\phi}
(g_{\alpha\beta}+\mathcal{U}_\alpha
\mathcal{U}_\beta) - e^{2\phi}\mathcal{U}_\alpha
\mathcal{U}_\beta,
\end{equation}
where $\phi$ is a new scalar field and $\mathcal{U}_\alpha$ is a \emph{constant} unit vector field directed along the time flow.  With AQUAL dynamics for $\phi$ Sanders' ``statified''  theory not only recovers the MOND phenomenology, but  also give lensing of the right strength~[21].   Of course, a constant vector field runs counter to the spirit of covariance, so the  theory is still not a consistent relativistic MOND.

For years the repeated failures to produce such a consistent theory had fuelled opposition to MOND by DM pundits.  The introduction of T$e$V$e\,S$~[22] ended that phase of the confrontation by exhibiting an example of a consistent relativistic implementation of MOND.  Today there are already quite a few gravity theories which claim to make the MOND paradigm relativistic.  Let me focus on T$e$V$e\,$S in a slightly more sophisticated form than it took at its inception.

\section{T$e$V$e\,$S, its structure and limits}
\label{sec:TeVeS}
Like RAQUAL or Sanders stratified theory, T$e$V$e\,$S has two metrics, the Einstein metric $g_{\alpha\beta}$ and the physical metric $\tilde g_{\alpha\beta}$, related via equation~(\ref{gg}), but with $\mathcal{U}_\alpha$ now regarded as dynamical.  The gravitational action (written exclusively with $g_{\alpha\beta}$, which also serves to raise indices within it) is the sum of three parts.  First, the usual Einstein-Hilbert action.  Second, the scalar's action
\begin{equation}
\label{Sscalar}
S_s=-\frac{1}{2 k^2 \ell^2 G}\int
\mathcal{F}\Big(k
\ell^2 h^{\alpha\beta}\phi_{,\,\alpha}\phi_{,\,\beta}\Big)\,(-g)^{1/2}\,d^4x;\qquad h^{\alpha\beta}\equiv g^{\alpha\beta}-g^{\alpha\mu} g^{\beta\nu}\mathcal{U}_\mu \mathcal{U}_\nu,
\end{equation}
wherein $\mathcal{F}$ is a positive function,  $k$ is a dimensionless coupling constant, and $\ell$ is a constant scale of length.  And thirdly the vector's action
\begin{equation}
\label{Svector}
 S_v =-{1\over 32\pi G}\int
\big[K g^{\alpha\beta}g^{\mu\nu} 
\mathcal{U}_{[\alpha,\mu]} \mathcal{U}_{[\beta,\nu]}+\bar{K} (g^{\alpha\beta} \mathcal{U}_{\alpha;\beta })^2-2\lambda(g^{\mu\nu}\mathcal{U}_\mu
\mathcal{U}_\nu +1)
\big](-g)^{1/2} d^4 x,
\end{equation}
where the square bracket denotes antisymmetrisation.  The vector's action sports a pair of dimensionless coupling constants, $K$ and $\bar K$ and a Lagrange multiplier field $\lambda$ charged with enforcing normalization of $\mathcal{U}_\alpha$ to -1.  Originally~[22] $S_v$ comprised only the $K$ term.  Contaldi et al.~[23] criticized T$e$V$e\,$S for the tendency of $\mathcal{U}_\alpha$ to form caustics, and suggested addition of the $\bar K$ term as a remedy; this was first used by Skordis~[24] and Sagi~[25].  Actually there are two more terms that are quadratic in derivatives of $\mathcal{U}_\alpha$ and have also been invoked (the set of four terms is often referred to as Einstein-Aether dynamics~[26]).  For simplicity I leave the last two terms out. 

Add now the standard action for matter and nongravitational radiation.  It is written exclusively with $\tilde g_{\alpha\beta}$ in order to comply with the weak equivalence principle, and to set T$e$V$e\,$S apart from standard GR.  In practice the consequent energy-momentum tensor $\tilde{T}_{\alpha\beta}$  is taken as a sum of perfect fluid contributions, one for dust-like matter and one for radiation.

Zlosnik, et al.~[27] have shown how to reformulate T$e$V$e\,$S  solely in terms of the metric $\tilde g_{\alpha\beta}$ (which turns out to have an Einstein-Hilbert action), and the vector field; the $\phi$ is eliminated with the help of the unit vector constraint.   In this form  T$e$V$e\,$S  might figuratively be described as GR with a DM (the vector field).  This form of the theory is useful for clarifying certain conceptual issues, but because of the intricate form of the vector action, it is much harder to calculate with.  So I continue to discuss T$e$V$e\,$S in its bimetric form.

In the limit $K\to 0$, $\bar K\to 0$ and $\ell\to\infty$ with $k\sim \ell^{-2/3}$,  T$e$V$e\,$S tends to GR for quasistatic systems and homogeneous cosmology~[22].  It is likely that it does so for more general situations.  It is thus advisable to keep $K, \bar K$ and $k$ small compared to unity in order not to stray too far from GR which, after all, is well tested in the solar system.    The above does \emph{not} mean that  T$e$V$e\,$S at large accelerations automatically becomes GR.   

To define ``large accelerations'' one must commit to a particular $\mathcal{F}$: every choice of $\mathcal{F}$  defines a separate  T$e$V$e\,$S theory.  I suggested a toy $\mathcal{F}$ in my original paper~[22].  Alternatives have been suggested by Sanders~[28] and by Angus, et al.~[29], the latter subsuming as special cases my toy $\mathcal{F}$ and that corresponding to the simple $\tilde\mu$, equation~(\ref{simple}).   One then derives the scalar equation 
 \begin{equation}
\left[\mu(k \ell^2 h^{\gamma\delta} \phi_{,\,\gamma}\phi_{,\,\delta})
h^{\alpha\beta}\phi_{,\,\alpha}\right]_{;\,\beta}
=kG\left[g^{\alpha\beta}+\left(1+e^{-4\phi}\right)\mathcal{U}^\alpha
\mathcal{U}^\beta\right] \tilde{T}_{\alpha\beta};\qquad \mu(y)\equiv\mathcal{F}'(y). \label{scalar_eq}
\end{equation}
Recalling that in the nonrelativistic limit the trace $\tilde g^{\alpha\beta}\tilde T_{\alpha\beta}$ tends to $-\rho$, while $ \mathcal{U}^\alpha \mathcal{U}^\beta T_{\alpha\beta}$ stands for energy density measured by an observer with 4-velocity $\mathcal{U}^\alpha$, it is clear that in the quasistatic nonrelativistic regime this last equation is very much like equation~(\ref{mod-Poisson}).

Now by linearizing the relation (\ref{gg}) using a first approximation to $\mathcal{U}^{\alpha}$, one concludes that $\Phi\approx \Phi_N+\phi$.  Thus, at least for quasispherical situations,  T$e$V$e\,$S reduces to AQUAL with
\begin{equation}
\tilde \mu\approx (1+k/4\pi\mu)^{-1}.
\end{equation}
The $\mathcal{F}$ must be chosen so that  this $\tilde \mu$ becomes linear in $|\nabla\Phi|$ at small gradients (small accelerations).  For example, with my toy $\mathcal{F}$  Milgrom's constant comes out as
\begin{equation}
a_0 = \sqrt{3 k}/4\pi\ell.
\end{equation}

For weak fields ($|\Phi|\ll1$) and large accelerations  ($|\nabla\Phi|\gg a_0$) the $\tilde\mu$ must approach unity so that the theory's predictions tend to those of Newtonian gravity.  This is indeed the case for my toy $\mathcal{F}$ and for Sanders' choice of $\mu$. Case $\alpha=1$  of Angus, et al.'s interpolating $\mu$ function  fails: it gives a vanishing effective Newton constant.  For a good $\mathcal{F}$ the Newtonian limit is gotten  by taking $\ell\to\infty$ in the weak field regime; one does not have to take $K\to 0$ and $\bar K\to 0$ which would be required to get GR.  One may thus envisage tests of T$e$V$e\,$S vs GR in strong gravity.  For example, Lasky, Giannios and Sotani in various combinations~[30,31] have considered the differences obtaining in neutron stars and black holes between GR and T$e$V$e\,$S, the latter sometimes generalized to include in the action all four terms quadratic in $\mathcal{U}_{,\alpha}$.

For weak fields and large accelerations one has to distinguish between the Newtonian and post-Newtonian behaviours.  The $\beta$ and $\gamma$ parametrized post-Newtonian (PPN) coefficients  of even the simplest  T$e$V$e\,$S ($\bar K=0$) agree with those of GR~[22,31].  The four $\zeta$ ``conservation coefficients'', Whitehead's  $\xi$ and the preferred frame coefficient $\alpha_3$ of generalized T$e$V$e\,$S all vanish (in harmony with GR) with no special demands on the theory's parameters.  Finally, the preferred frame coefficients $\alpha_1, \alpha_2$, though nonvanishing, can be made to comply with the stringent solar system experimental constraints in appropriate ranges of the four $K$ and single $k$ coupling coefficients of fully generalized T$e$V$e\,$S~[25].
 T$e$V$e\,$S thus squeaks through the PPN filter; however, its recovery of the Newtonian limit leaves much to be desired.  It is true that  as $|\nabla\Phi|/a_0$ grows well beyond unity, $\nabla\Phi\to\nabla\Phi_N$, but the approach is not as fast as one would like.  Consequently the Kepler constant $a^3 P^{-2}$ is not exactly constant from planet to planet as would be required by Kepler's third law.  This prediction may already be in conflict with modern ephemerides, and the problem may not be solvable within the general framework of tensor-vector-scalar type theories~[28].

Thus T$e$V$e\,$S (and theories like it) may not be the final word.   But how does the theory fare in light of galaxy phenomenology, gravitational lensing and cosmological data?  On the answers hinges the decision whether one needs to look elsewhere for a panacea alternative to DM.  

\section{Gravitational lensing and cosmology in TeVeS}
\label{sec:galaxies}

 In the \emph{nonrelativistic} regime T$e$V$e\,$S gives back AQUAL$/$MOND (\S\ref{sec:TeVeS}).  This is shown rigorously for spherical systems~[22], but apparently also goes through in the presence of considerable asymmetry~[29].  Turning to relativistic issues, recall that in RAQUAL (with either conformal or disformal relations between the two metrics), the deflection of light is hardly different from that which GR would predict with no DM---in conflict with observations.  By contrast in T$e$V$e\,$S the (single) potential responsible for weak field light deflection is the same as that which appears in AQUAL$/$MOND~[22].  Since this last describes the dynamics of a galaxy as well as an optimal DM halo model (it produces the same potential:  \S\ref{sec:paradigm}), its predictions for lensing will thus be comparable to those of GR with an optimal distribution of DM.  

However, this blanket statement has to be qualified.  AQUAL$/$MOND with the customary $\tilde\mu$ predicts that a galaxy's RC must remain flat to arbitrary large radius; the logarithmic potential $\Phi$ that accomplishes this will induce light bending of the form caused by an isothermal sphere in standard gravity theory.  By contrast, the limited extent of DM halos should cause RC's to droop at large distances, and force the light bending at  large impact parameter to approach that of a point mass.  The Keplerian descent of the RCs is not seen.  But since gravitational lensing can probe a galaxy's potential to larger radii than can radio and optical Doppler measurements, do gravitational lenses resolve the above confrontation?   

Microlensing in our galaxy aside, gravitational lenses and their objects lie at cosmological distances.  Thus theoretical predictions for lensing \emph{a la} T$e$V$e\,$S must use a genuine T$e$V$e\,$S cosmological model for consistency.   This was first done by Chiu et al.~[32] and Zhao et al.~[33].  In line with the early finding  that T$e$V$e\,$S isotropic models with baryonic matter content alone resemble the corresponding GR models because the scalar $\phi$'s energy density stays small~[22], these workers find the former to provide just as consistent a scaffolding  for the analysis of cosmological gravitational lenses as do isotropic GR models.  

Chiu et al.~[32] derived the T$e$V$e\,$S lensing equation and applied it to strong lensing (multiple images).  Most gravitational lenses are giant elliptical galaxies.  Chiu et al. modelled the baryonic mass distribution of such a galaxy by a point; an  Hernquist mass model is probably better.  Zhao et al. employed both in their comparison of T$e$V$e\,$S's  predictions with a sample of quasars doubly imaged by intervening galaxies from the CASTLES catalogue.    They estimated the lens mass in each case by comparing predicted and observed image positions or amplification ratios; the two methods gave consistent results.  The derived mass-to-light ratios fall in the normal range for stellar populations, with some outliers.  Shan et al.~[34], who took departures from spherical symmetry into account,  also found agreement with T$e$V$e\,$S  for those in their sample of 15 doubly imagined quasars with lenses not situated in clusters (to exclude the external field effect).

The opposite conclusion was urged in a trilogy of papers by Sakellariadou and coworkers [35].  They concluded that  T$e$V$e\,$S will not work for galaxy lenses in the CASTLES catalogue without help from invisible matter.  In the first paper Ferreras et al. actually calculated lensing using a mixture of MOND and GR instead of  T$e$V$e\,$S, a procedure known to be misleading [15, 32].  In the second attempt, Mavromatos et al. reiterate the above claim based on comparison of their T$e$V$e\,$S mass estimates for 18 lenses in CASTLES with the masses inferred from luminosity.  But recently Chiu et al. [36], who carefully repeated the comparison for 10 of the lenses in question which can be modelled as spherical, reached the \emph{opposite} conclusion that with the $\tilde\mu$ deriving from the toy $\mathcal{F}$ or the simple $\tilde\mu$, T$e$V$e\,$S works adequately without requiring invisible mass.  They impute the conclusion of Sakellariadou and coworkers to improper comparison of the inferred lensing mass with the stellar mass implied by the optical intensity through a fixed aperture.

The distribution by angular separation of the two images in lensed quasars has proved troublesome for the DM paradigm.  In T$e$V$e\,$S it has been investigated by Chen \& Zhao~[37] and lately by Chen~[38].   Again modelling the mostly elliptical galaxies with Hernquist profiles, and describing their space distribution with the Fontana function, these workers compare predictions of, first T$e$V$e\,$S for a purely baryonic universe with cosmological constant, and second GR with DM and baryons, with the observed statistics in the CLASS/JVAS  quasar survey.   They report that  T$e$V$e\,$S comes out on top.  The above is for spherical mass models of the galaxies; a step towards the modelling of asymmetric lenses within  T$e$V$e\,$S has been taken by Shan et al.~[34].

When the colliding clusters or ``bullet cluster'' system 1E0657--56 was discovered~[39], it was proclaimed as a ``proof of dark matter''. Weak lensing mapping (of the distortions of background galaxies) seems to show the main gravity sources to lie with the galaxies, rather than in the ejected hot gas which dominates the visible baryonic mass.  Collisionless DM would indeed be expected to get separated from the dissipational gas and go along with the galaxies; hence the GR based conclusion that much DM exists in 1E0657--56.   Angus et al. ~[40] analysed 1E0657--56 within T$e$V$e\,$S and concluded that Sanders' suggestion~[41] that clusters contain, along with the baryons, much mass in few-eV neutrinos  might just save the day.   Confirming that the source of gravity in 1E0657--56 must include an invisible component, Feix et al.~[42] assert that nonlinearity of the AQUAL equation cannot prevent the lensing from tracking the baryonic matter (just the opposite is seen).  But as Ferreira \& Starkman remark~[43],   the absence in T$e$V$e$S of a Birkhoff-type theorem allows nearby structures, e.g. nearby clusters, to influence 1E0657--56's gravitational field.  This effect has not been factored into any of the above arguments against T$e$V$e$S.

It is not the \emph{asymmetry} of the ``bullet'' system which is  the reason for  T$e$V$e\,$S's woes.  Takahashi \& Chiba~[44] analysed observed weak lensing by three quasispherical clusters.  For each they obtained the gravitational field with TeVeS using several acceptable $\tilde\mu$ functions and a model mass profile inferred from a large cluster sample from the SDSS survey. Their predicted shear and convergence of the lensed light failed to fit the observations unless they added an unrealistically large mass in neutrinos.   Similar conclusions are reached by Natarajan \& Zhao~[45].    In retrospect it is no wonder that  T$e$V$e\,$S has problems with cluster lensing.  After all, MOND dynamical models of clusters do not clear all the missing mass problem, and T$e$V$e\,$S is built  on a MOND foundation. 

Obviously T$e$V$e\,$S's performance with clusters is weaker than that with galaxies.  How does it do as we go to the domain of large scale?
Turn to cosmology.  The Friedmann models of GR can be transplanted to T$e$V$e\,$S~[22].  One can ask whether $G$ and $a_0$ are really constants in T$e$V$e\,$S cosmology.  Whereas scalar tensor theories, e.g. Brans-Dicke's, generically have a variable effective gravitational constant,   $G$ is strictly constant in T$e$V$e\,$S isotropic cosmology~[46].  By contrast, $a_0$ varies, but does so on a timescale much longer than Hubble's~[46], at variance with some early speculations on MOND~[6].  Since $a_0$ can be determined from a disk galaxy's RC, it should be possible to check this consequence of T$e$V$e\,$S with RC's at $z\sim 2-5$; these RC's will be available within years.

Friedmann models are overly idealized.   In his exemplary paper Skordis~[47] worked out the full covariant formalism for growth of cosmological perturbations in T$e$V$e\,$S. Skordis et al.~[48] then used it to show that T$e$V$e\,$S's predictions can  be made consistent with the power spectrum of the galaxy distribution and that of the CMB fluctuations, if one allows for a contribution of massive neutrinos  to the tune of $\Omega_\nu=0.17 $, as well as cosmological constant, but no standard DM.  The role of DM is here taken over by a feedback mechanism originating in the scalar field~[47] (or in the vector field according to a variant opinion~[49]).   Thus  power spectra do not cleanly distinguish T$e$V$e\,$S from GR.

Dodelson and associates~[50,51] pointed out that different modified gravity theories lead to relations between the matter overdensity and the local depth of the cosmological gravitational potentials, relations  different from the standard GR ones.  These are reflected in different integrated Sachs-Wolfe signals for each theory which  can be brought out by examining the correlation between CMB temperature fluctuations and galaxy number overdensity~[50].  Alternatively, the different relations are reflected in different correlations between gravitational lensing convergence (and shear) and galaxy number overdensity~[51]. Both of the above procedures are confused by biasing (the unknown ratio of matter to galaxies clustering).   Zhang et al. have proposed an estimator, $E_G$, which should make it possible to factor out biasing while correlating fluctuations in galaxy number overdensity with lensing~[51].   They conclude that planned surveys in the optical by the LSST,  or in the radio by the SKA, will reach the precision requisite to distinguish TeVeS from GR.

Recently Reyes, et al.~[52] estimated $E_G$ from a sample of $7\times 10^4$ giant ellipticals with a mean $z=0.32$, and claimed a statistically significant difference between the predicted $E_G$ for a T$e$V$e\,$S' model and the measured one, whereas GR's $E_G$ fit well.  The T$e$V$e\,$S cosmological model in question~[51] comes from T$e$V$e\,$S with only the $K$ term in vector action~(\ref{Svector}), the version known to develop caustics~[23] and not to fit all the measured PPN coefficients~[25].  And the cited prediction for $E_G$ is actually for a redshift interval larger than the observed one (T$e$V$e\,$S's and GR's predictions for $E_G$ converge at low redshift~[51]).   Finally Reyes et al.'s error bars are large; a much larger sample is required for a statistically significant discrimination between the theories~[51].  The claim of Reyes et al. paper's title to have established GR is surely premature.

I thank Joao Magueijo for useful remarks, and the organizers for the invitation.

\label{lastpage} 
\end{document}